# Recent Increases in Tropical Cyclone Rapid Intensification Events in Global Coastal Regions

Yi Li[1,2,3], Youmin Tang[3,4,5]*, Shuai Wang[6], Ralf Toumi[7], Xiangzhou Song[1,2,3]

[1] Key Laboratory of Marine Hazards Forecasting, Ministry of Natural Resources, Hohai University, Nanjing, China

[2] Key Laboratory of Ministry of Education for Coastal Disaster and Protection, Hohai University, Nanjing, China

[3] College of Oceanography, Hohai University, Nanjing, China

[4] University of Northern British Columbia, Prince George, Canada

[5] Southern Marine Science and Engineering Guangdong Laboratory (Zhuhai), Zhuhai, China

[6] Program in Atmospheric and Oceanic Sciences, Princeton University, Princeton, NJ, USA

[7] Department of Physics, Imperial College London, London, SW7 2AZ, UK

*Corresponding author: Youmin Tang, ytang@unbc.ca

# Abstract

Rapid intensification (RI) is likely the most crucial contributor to the development of strong tropical cyclones and the largest source of prediction error resulting in great threats to life and property, which can become more threatening with proximity to landfall. While enormous efforts have been devoted to studying the basin-wide fluctuation, temporal-spatial variations of global RI events remain uncertain. Here, we show that, compared with open oceans where the annual RI counts do not show any significant change, the coastal offshore regions within 400 km from the coastline host significantly more RI events, with the RI count tripled from 1980 to 2020. Reasons responsible for the coastal RI occurrence are analysed, with the dominant large-scale environmental factors identified. This work yields an important new finding that an increasing threat of RI in coastal regions has occurred in the preceding decades, which may continue in a future warming climate.

Fluctuations in tropical cyclone (TC) activity are a major concern in densely populated coastal regions[1,2]. Although the frequency of TCs has been declining in observational data and numerical projections[3–5], the threat of intense TCs has increased[6–10] continuously. Hence, accurate predictions are required to improve preparedness and reduce TC-related damages to life and property. Unfortunately, despite improvements in TC track forecasting, errors in intensity forecasting have not significantly decreased over the past decades[11,12]. Rapid intensification (RI) occurs when a TC intensifies dramatically over a short period, rendering its forecasting particularly challenging owing to the uncertainty in onset time and duration[13–16]. RI also affects the development of some of the most intense TCs[17], as recently demonstrated by Cyclone Nagis (2008)[18], Typhoon Hato (2017)[19] and Hurricane Harvey (2017)[20], all of which caused catastrophic damage. Therefore, accurate prediction of RI is paramount, especially in vulnerable coastal regions[16].

Previous studies have indicated that the intensification rate of major TCs is increasing in different ocean basins[21,22], and the number of TCs undergoing RI is also rising[23,24]. This warrants more attention as no clear trends have been defined despite a direct threat posed by RI in coastal regions. This study analysed the trends in the counts of RI events across global oceans from 1980 to 2020, as the TC measurements obtained in this period are reliable because of the wide use of satellite observations and post-season analysis[25,26]. In addition, global surface temperatures have also increased at a rate of ~0.18 °C per decade since 1981, which is more than twice of that since 1880 (0.08 °C)[27]. In a warming climate, changes in vertical wind shear, mid-level humidity, and ocean temperature are all likely to affect RI and other TC properties[3,28]. At regional scales, the oceanic and atmospheric conditions changes are non-uniform. The faster warming in the western boundary currents[29], for example, might favour TC intensification in these regions[30].

Aside from the ambient environmental vertical wind shear (VWS)[31] and relative humidity (RH)[32] that govern TC intensification, the upper ocean also plays an important role in fuelling the overlying atmosphere and TC intensification, as high ocean temperatures favour TC development by increasing the thermodynamic potential intensity (PI)[33,34]. The difference between PI and the current intensity of a TC, or its potential intensification rate (PIR), must be large for its development and RI[35].

In this study, the trends in RI and environmental conditions are analysed based on the International Best Track Archive for Climate Stewardship (IBTrACS) dataset[36] and the fifth generation of the European Centre for Medium-Range Weather Forecasts (ECMWF) reanalysis (ERA5)[37]. RI is the increase in the maximum sustained surface wind speed by at least a certain threshold within 24 h. The 45 kt/24 h threshold was used according to recent derivations via objective joint clustering[38]. The results were compared with those obtained using other thresholds, including the more widely used 30 kt/24 h[13]. To the best of our knowledge, this study is the first to show the increasing trend in RI events in coastal regions on a global scale and identify the dominant environmental factors.

## Results

## Trends in RI

The trends in the counts of RI events within 400 km from the coast were first analysed. Given the translation speed is approximately 4–5 m/s (350–430 km/day)[28], a TC entering regions within 400 km from the coast would likely make landfall within one day; hence, this becomes an urgent concern for operational forecast owing to the relatively short amount of time for analysis. The annual number of RI events in these coastal regions between 1980 to 2020 increased by 3.0 ± 0.8 per decade (mean ± standard deviation, Fig. 1A). Over the coastal waters, less than five RI events occurred per year in the 1980s, whereas the annual count increased to approximately 15 by 2000. In addition, the time fraction of RI periods within the lifespan of a TC has also constantly increased, in which a 2.5 ± 0.6% increase per decade was recorded during this 41-year period (Fig. 1B). Similarly, the ratio of coastal occurrences of global RI events has increased by 3.2 ± 1.8% per decade (Fig. 1C). Similar trends were observed using the conventional threshold of 30 kt/24 h (Fig. S1) and the annual number of RI events increased by 5.1 ± 1.8 (13.6 ± 4.9%) per decade ($p < 0.01$), with the time fraction also increasing by 4.1 ± 1.2% per decade.

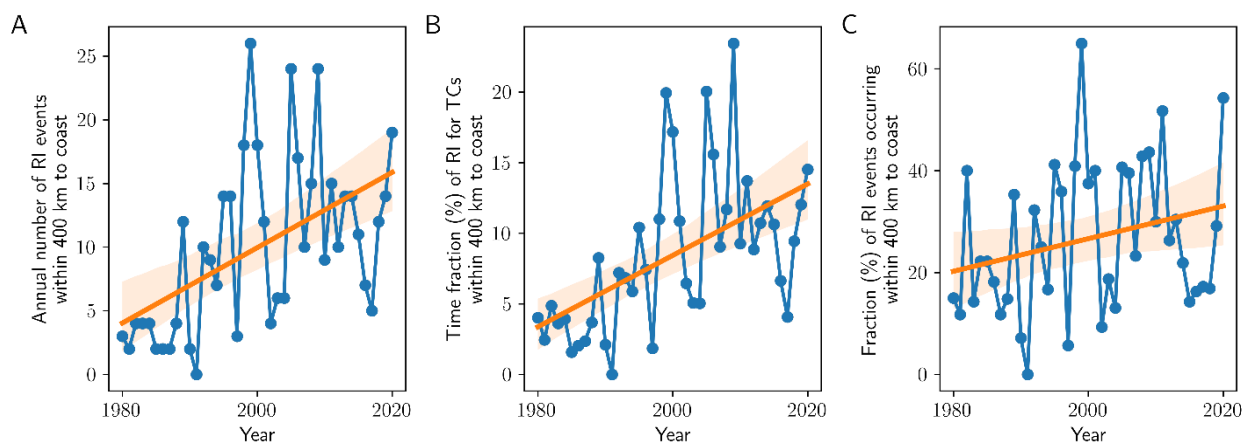

**Fig. 1. Time series of global RI activity in the coastal regions.** **A** Annual mean count of RI events within 400 km from the coast. **B** Annual mean time fraction of RI for TCs within 400 km from the coast. **C** Ratio of global RI events occurring within 400 km from the coast. RI is defined as an intensification of at least 45 kt/24 h. The blue lines and dots show historical data. The orange lines show linear trends, with shading denoting a 95% confidence interval.

The variability of the annual count of RI events as a function of the distance to land was then calculated (Fig. 2A). The most rapid increase occurred at 200 to 400 km from the coast, where the annual number of RI increased by 1.9 ± 0.6 per decade ($p < 0.01$). The trend reduced almost linearly with proximity to land by 0.4 ± 0.2 RI event/decade/1000 km ($p = 0.04$). As the range extended to 2,000–2,200 km, the annual number of RI events slightly increased by 0.3 ± 0.2/decade ($p = 0.18$). As the average annual RI count was 1.6 for each 200-km bin, this distance-dependent variation can reach up to 15%/decade/1,000 km or higher. Consistent results were obtained using different thresholds of minimum landmass sizes such as 2,000, 5,000, and 10,000 km$^2$ (Fig.

S2). When different thresholds of RI such as 30 kt/24 h and 40 kt/24 h were used, evident landward shifts were also observed (Fig. S3). When the commonly defined RI with an intensification of at least 30 kt/24 h was used, there was an increase in RI events by 1.1 ± 0.3/decade, accompanied by a decrease in the distance to land by 1,000 km. The results obtained from the ADT-HURSAT data also showed similar patterns (Fig. S4), indicating a valid global-scale landward variation of RI.

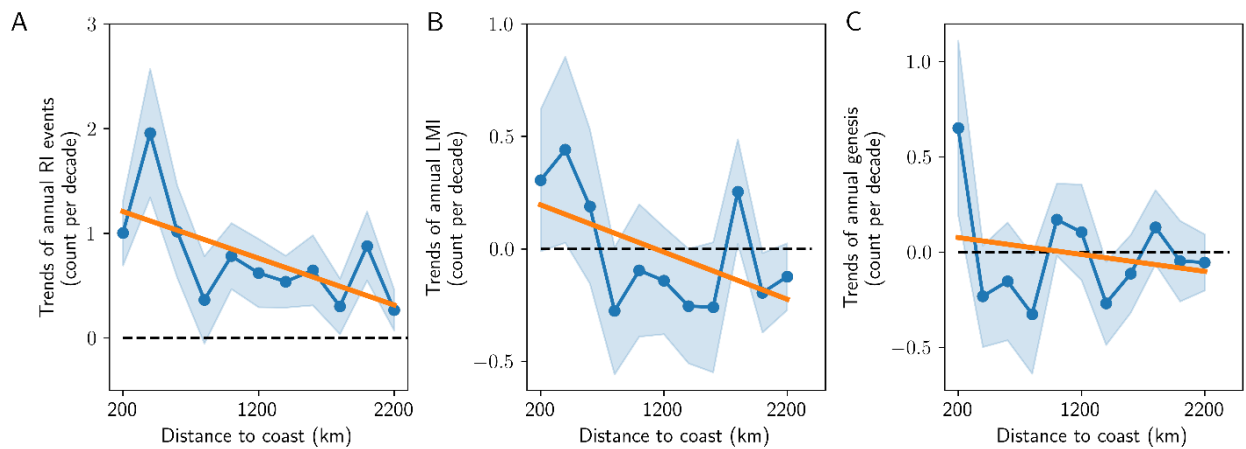

**Fig. 2. Linear trend of annual counts of RI, LMI, and genesis with different distance-to-land.** **A** Trends of annual count of RI events. **B** Trends of annual life maximum intensity (LMI) of TCs. **C** Trends of the annual genesis of TCs. RI is defined as an intensification of at least 45 kt/24 h. The x-axis is the distance-to-land from 0–200 km to 2,000–2,200 km, with a 200-km interval. The blue lines and shadings show linear temporal trends and a 95% confidence level of the trends, respectively. The orange lines show linear fits of the temporal trends as a function of distance-to-land.

To examine the robustness of the observed landward variation, or the notable increase in RI events landward, data from the pre-satellite period (1951–1979) were investigated. Although fewer RI events (85.9/year) occurred during this period compared with those in the satellite period (139.6/ year), the same landward variation was apparent, with magnitudes of 0.3 to 1.3/decade/1,000 km, depending on the selection of RI threshold (Fig. S5).

In addition, locations of RI are also associated with the locations of other TC activities such as the genesis and lifetime maximum intensity (LMI). Landward migrations towards the coast were found in both the TC activities, but were not significant ($p$ values are 0.09 and 0.5, respectively, Fig. 2B and C). Moreover, the trends were substantially smaller for both genesis and LMI (0.2 and 0.09/decade/1,000 km, respectively) than that of RI (0.4/decade/1,000 km), indicating a stronger coastward variation of RI.

TC-related damages are dependent on the population and economy exposed. The western North Pacific, Bay of Bengal, Madagascar, Mozambique, Caribbean, and the Gulf of Mexico are considered the most vulnerable regions[1]. Significant increases in RI counts were observed in most of these regions (Fig. 3A), especially in Mozambique, the Philippine archipelago, and Central America. Specifically, RI events here can increase by up to 0.07 (2.5%)/decade for each 2° × 2° latitude-longitude grid. Meanwhile, significant declines were observed in the Central Atlantic and Pacific. The landward variation occurred regardless of the RI threshold. When the threshold of 30 kt/24 h was used, similar spatial distributions were observed, except for a patch of decrease located in the Eastern Pacific (Fig. S6). The 95[th] percentile of 24-h intensity changes also showed similar patterns based on the calculations obtained from the Monte Carlo experiment. The distance-trend diagrams of individual basins verify the significant increases in RI in coastal regions, except in the Eastern Pacific where increases also occurred over the open ocean, and North Indian Ocean where the most rapid increase was observed 600–800 km from the coast (Fig. S7).

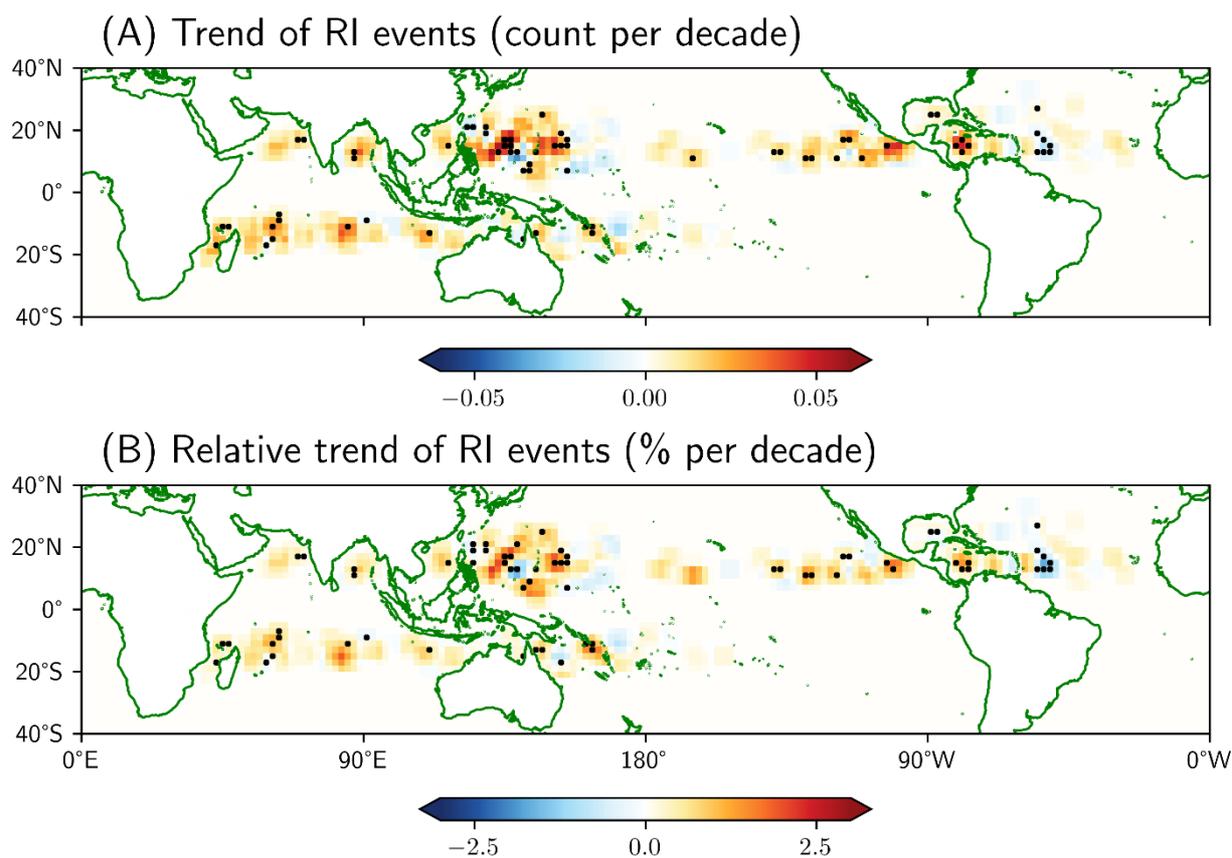

**Fig. 3. Spatial distribution of the linear trend of annual RI counts. A** Trend in the annual count of RI events. **B** Relative trend in the annual count of RI events. The counts and ratios were calculated for each 2 × 2 latitude-longitude

grid. Black dots show areas where 95% confidence for the linear fit was satisfied. Data smoothing using a three-point smoother was performed for better display clarity.

**Influences of large-scale environmental factors**

As long-term variations in RI are potentially influenced by large-scale circulation and global ocean conditions, the effects of internal climate variability and global warming were investigated. The internal climate variability was measured using climate models such as El Niño–Southern Oscillation (ENSO) and Pacific Decadal Oscillation (PDO), while the effects of global warming were estimated using the trend of global mean sea surface temperature (SST) between 60 °S and 75 °N[39]. The impacts of these metrics were eliminated by linearly regressing the time series of the RI onto the climate indices[25]. Marginal differences were found as the Niño3 index was removed, and the amplitude of RI landward variation (Fig. 4A) maintained at 0.4 ± 0.2/decade/1,000 km. This indicates that the impacts of ENSO on these long-term landward variations are negligible, although it may have a substantial role in inter-annual variability, hence further exploration is required. Meanwhile, the PDO or global mean SST alone moderately influenced RI. With the removal of their impacts, the landward variation rates decreased slightly to 0.3/decade/1,000 km for PDO and 0.2 for SST (Fig. 4B and C). However, it should be mentioned that without the impacts of the global SST, the increasing rate of RI events declined to approximately 0 per decade, particularly in regions that were beyond 600 km from the coast. This suggests that global warming is a key driver of increases in RI events; however, the impacts are spatially non-uniform and limited within the open ocean. In contrast, when both the PDO and global SST trend were removed, the landward variation of RI decreased dramatically to 0.05 ± 0.0/decade/1000 km ($p = 0.3$), indicating that the increasing trend is dominated by both these factors (Fig. 4D). Other metrics, including the ENSO-Modoki index[40] and North Atlantic oscillation, did not have a substantial influence on RI (Fig. S8).

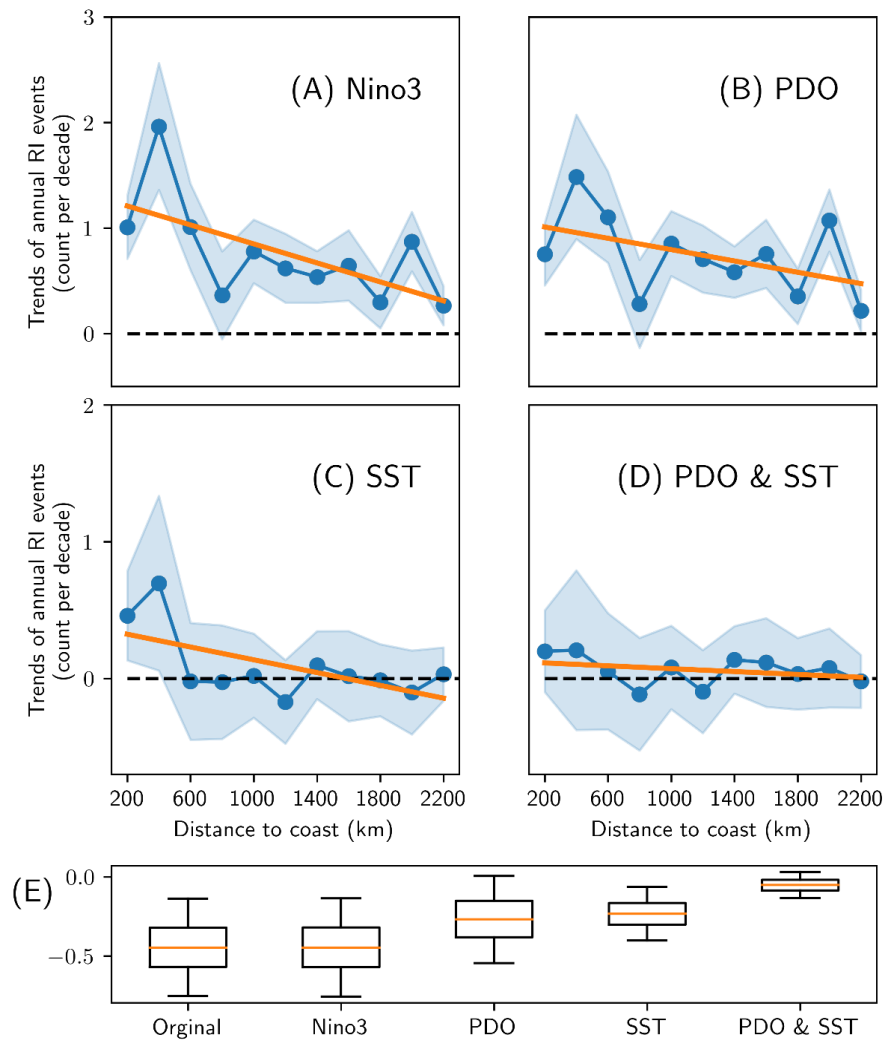

**Fig. 4. Linear Trend of RI events with climate indices and/or global SST trend reduced. A-D** Trend of the annual count of RI events with **A** Niño3 index, **B** PDO index, **C** global SST trend, and **D** both PDO and SST trend reduced. The x-axis is the distance to land from 0–200 km to 2,000–2,200 km, with a 200–km interval. The blue lines and shadings show linear temporal trends and a 95% confidence level of the trends, respectively. The orange lines show linear fits of the temporal trends as a function of distance to land. **E** Magnitudes of landward variation (i.e., slopes of orange lines in **A-D**). Orange line is the median, boxes extend from 25$^{th}$–75$^{th}$ percentiles, while whiskers show the 5$^{th}$ and 95$^{th}$ percentiles.

## Influences of the local environment

High intensification rates and RI formation typically involve weak VWS, high mid-level RH, and high PI. The trends of PI, calculated using either surface (maximum potential intensity, MPI) or subsurface ocean temperatures (ocean-coupling potential intensity, OCPI), showed spatial patterns similar to those of RI events (Fig. 5A and C). Substantial increases were found over the Western Pacific, South China Sea, Caribbean Sea, Gulf of Mexico, and Southern Indian Ocean, except in the Eastern Pacific where OCPI declined with an increase in MPIs. The increase in OCPI was more apparent than that of the MPI over coastal waters, and the linear regression coefficients as a function of distance-to-land ratio were 0.7 and 1.6 kt/decade/1,000 km, respectively (Fig. 5B and D).

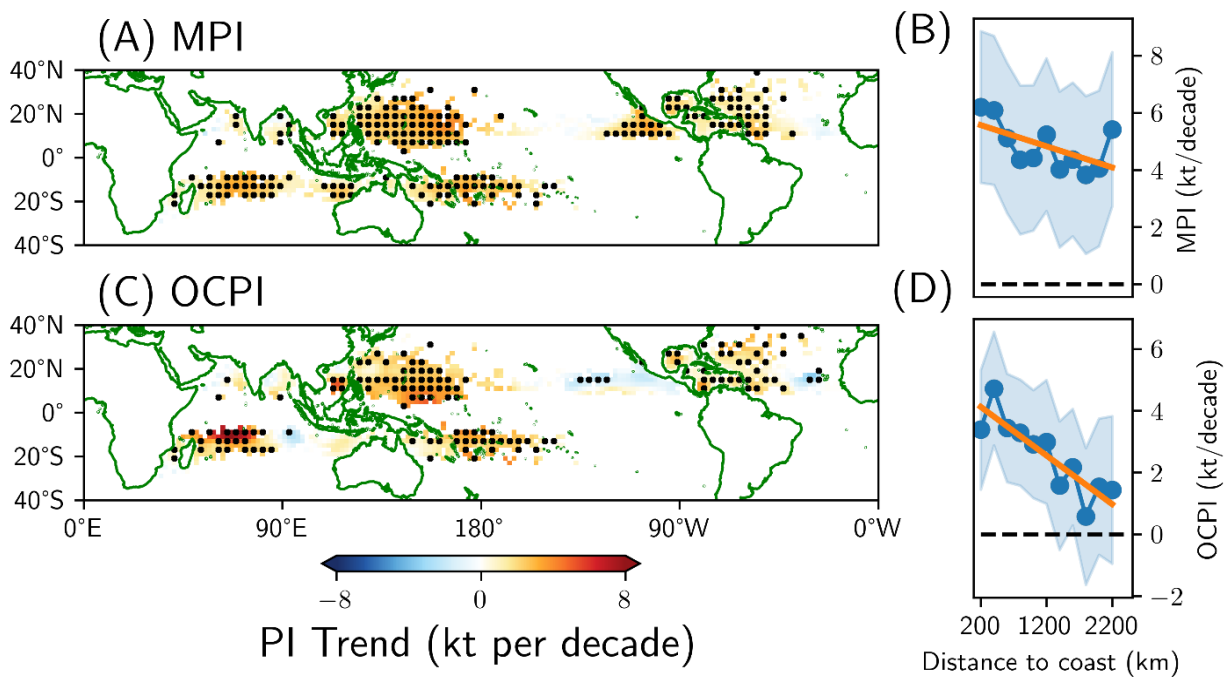

**Fig. 5. Linear trend of potential intensity. A** Spatial distribution of the linear trend of maximum potential intensity (MPI). **B** Linear trend of MPI with different distances-to-land. **C** Spatial distribution of the linear trend of ocean-coupling potential intensity (OCPI). **D** Linear trend of MPI with different distances-to-land.

Two potential intensification rate[35] metrics, namely MPIR and OCPIR, were further analysed by computing the difference between the current intensity and MPI or OCPI. Both metrics generally follow the distribution of MPI and OCPI, and the increase can reach up to 10 kt/24 h/decade (Fig. S9). However, both these metrics significantly decreased in the mid-western North Pacific, where decreases in RI were also found.

Globally, the MPIR and OCPIR over coastal waters increased more rapidly (3.7 kt/24 h/decade/1,000 km) compared with that over open ocean (4.1 kt/24 h/decade/1,000 km). As the average MPIR and OCPIR were 170 and 108 kt/24 h, respectively, the increases were ~6–10%/decade and the landward variation was ~2–4%/decade/1,000 km.

Other RI-related environmental factors including VWS and mid-level RH were also analysed (Fig. S10). The wind shear declined significantly in some regions where RI increased: the Indian Ocean, South China Sea, and Eastern Pacific. For the global mean value, the wind shear weakened within 800 km from the coast and increased by 0.17 m/s/decade/1,000 km towards the ocean. These oceanward variations in wind shear favour the development of RI in coastal regions. An increase of 0.17 %/decade/1,000 km was also detected over coastal waters for mid-level humidity. However, it was primarily significantin the North Indian Ocean, suggesting that this factor may not be important for other basins.

## Discussion

Forecasting RI is of utmost importance for the prediction and preparedness of TC and remains a major challenge in coastal regions. In this study, we analysed TC observations from 1980 to 2020 and found that RI counts increased significantly more rapidly in coastal regions than in open oceans, in which the combination of the PDO and global ocean warming plays a significant role in this landward variation. These increases are attributed to the potential intensity driven by the ocean temperature. However, ambient atmospheric conditions, including the vertical wind shear and mid-level humidity, initiated significant ($p < 0.05$) changes mainly in the Indian Ocean.

The linear trend analyses performed are prone to uncertainties, particularly regarding the data and the selected period. To address this, we compared the IBTrACS and ADT-HURSAT, and found similar landwards trends among both datasets. In addition, an improved least-squares algorithm was performed[41], which also detected high increases towards the coast (Fig. S11). The trend was consistent regardless of the use of the RI threshold (30, 35, 40, or 45 kt/24 h) and the minimum size of the landmass (1,400, 2,000, 5,000, and 10,000 km$^2$).

In a recent study, a landward migration of TC activities, such as LMI and time fraction, was observed due to the enhancement of the westerly steering flow[42]. In the present study, the genesis locations also showed landward movement, although the trend was not statistically significant. Although the locations of RI are closely associated with the movement of these activities, the former cannot be entirely explained by the latter, as their landward shifts are significantly weaker than those of RI. Hence, RI is possibly more influenced by internal processes and its local environment than by large-scale steering flow[43,44].

In this study, we focused on the RI trends during the past 40 years and the analyses were limited to this period due to the lack of quality of earlier best-track data. Linear regression analyses indicated that landward variation could not be solely attributed to either internal climate variability or global warming. Arguably, a 40-year-long period can be insufficient to demonstrate long-term trends or evidence of global warming[3]. Therefore, conducting further analyses using reliable numerical simulations is necessary. Previously, TC activities, including their genesis and tracks, have been examined using climate models and reanalyses[3,5,8]. However, the existing models are still unable to resolve RI locations and magnitudes directively[28], and therefore, models with higher resolution or advanced downscaling techniques are needed[45,46].

Considering the relatively slow progress on improving TC intensity forecasts, the increasing trend of RI in coastal regions poses greater challenges and concerns for operational forecasting. Although the RI trend in climate projections was not explicitly explored here, this study demonstrated the fundamental role of potential intensity in increased occurrences of RI events. The projected warming of coastal waters presents an environment with even higher potential intensity and, in turn, more RI events. The continuously increasing population and economy in coastal regions also indicate that these areas will potentially have higher exposure and vulnerability to TC threats. Therefore, prediction and preparation of TCs would be more challenging and further efforts on improving RI prediction are therefore required.

# Methods

## Tropical cyclone data

TC best-track observations were obtained from the International Best Track Archive for Climate Stewardship (IBTrACS, v4r00)[36] in which the data were provided by U.S. agencies, i.e., National Hurricane Center (NHC) and Joint Typhoon Warning Center (JTWC). Another dataset, ADT-HURSAT, was developed by processing the HURSAT satellite imagery using the Advanced Dvorak Technique (ADT)[47] and has been used in several trend analyses[21,48,49]. Here, we also examined ADT-HURSAT for the period from 1980 to 2017 and compared the results from the two datasets.

We only considered tropical cyclones (lifetime maximum intensity (LMI) ⩾ 64kt), over the period 1980–2020. In addition, to eliminate the influence of topographic effects and extra-tropical transition, we chose only TC tracks over the ocean and within the range of 40°S and 40°N. We only took the records at the standard observational times: 00, 06, 12, and 18 Coordinated Universal Time (UTC). This pre-processing was similar to that performed in the study of RI[13,50].

In IBTrACS, the default distance to the nearest land, including all continents and islands larger than 1,400 km$^2$ (equivalent to the area of Kauai, Hawaii), was provided for each best-track geographical location. We also examined other thresholds of minimum landmass, including 2,000, 5,000 and 10,000 km$^2$, using the coastline data obtained from Global Self-consistent, Hierarchical, High-resolution Geography (GSHHG) database[51].

## Environmental data

The monthly mean atmospheric data, including relative humidity (RH), wind, and atmospheric temperature, were obtained from the fifth generation of ECMWF reanalysis (ERA5) dataset[37], while the ocean temperature was provided by the Ocean ReAnalysis System 5 (ORAS5)[52]. The horizontal resolutions of the ERA5 and ORAS5 data were both 0.25°.

## Definitions of RI and genesis

The intensification rate was calculated as the change in the maximum surface wind speed ($V_{max}$) in 24 h. RI is commonly defined as an increase in the surface maximum wind speed ($V_{max}$) of at least a threshold within 24 h[13]. The threshold of 30 kt (15.4 m/s), as recommended by Kaplan and DeMaria[13], is widely used. However, other thresholds exist[17], and recently a physically robust value of 45 kt/24h was derived via objective joint clustering[38]. We used 45 kt/24 h as the major threshold and compared results using different thresholds. The genesis location was defined as the position at which a TC first reaches an intensity of 20 kt.

## Statistical information

In this study, we calculated the trend via linear regression. However, the robustness of linear regression is usually prone to the choice of period. We, therefore, implement an improved Ordinary Least-Square (OLS) algorithm[41] and compare the results with least-square linear regression. The details of this OLS algorithm are provided in the supplementary text.

The uncertainty of $V_{max}$ could be up to 10 kt[26]. Thus, we also performed Monte Carlo experiments to estimate the uncertainty of the 95$^{th}$ percentile of 24-h Vmax change. Following Bhatia et al.[21], 1,000 samples were produced by adding random noise from a uniform distribution on the interval ±10 kt to each $V_{max}$ change.

## Environmental variables

The vertical wind shear was calculated as the amplitude of wind vector difference between 200- and 850-hPa pressure levels. The mid-level relative humidity was obtained at 600-hPa.

Two potential intensity (PI) indices were used in this study, namely the maximum potential intensity (MPI)[53,54] and ocean coupling potential intensity (OCPI)[34]. MPI was calculated as a function of SST,

$$\text{MPI} = \sqrt{\frac{\text{SST}-T_o}{T_o}\frac{C_K}{C_D}(k^* - k)}, \qquad (4)$$

where $T_o$ is TC outflow temperature determined by the atmospheric vertical profile, $C_D$ the drag coefficient, $C_K$ enthalpy exchange coefficient, $k^*$ the saturation enthalpy of the sea surface, and k the surface enthalpy. OCPI is a function of the average ocean temperature of upper 80 m ($T_{80}$),

$$\text{OCPI} = \sqrt{\frac{T_{80}-T_o}{T_o}\frac{C_K}{C_D}(k^* - k)}. \qquad (5)$$

The potential intensification rate (PIR)[35], was calculated from PI.

$$\text{PIR} = \frac{C_D}{H}(EV_{PI}^2 - V^2), \qquad (6)$$

where $V_{PI}$ is the potential intensity (either MPI or OCPI), V the current TC intensity, H the height parameter and E the dynamical efficiency,

$$E = \left(\frac{f+\frac{2V}{r}}{f+\frac{2V_{PI}}{r}}\right)^n, \qquad (7)$$

where n is a sensitivity constant, r is the radius of maximum wind speed, and f is the Coriolis parameter. As suggested by Wang et al.[35], the parameter used were $C_D = 2.4\times10^{-3}$, $C_K=1.2\times10^{-3}$, H = 3 km, n = 1, and r = 15 km.

## Data Availability

The TC best track data used in this study were IBTrACS Version 4 (v4r00) retrieved from NOAA National Centers for Environmental Information https://www.ncei.noaa.gov/products/international-best-track-archive, and ADT-HURSAT from Kossin et al.[5] ERA5 and ORAS5 data were downloaded from Copernicus Climate Data Store https://cds.climate.copernicus.eu/#!/home. GSHHG Version 2.3.7 was obtained from Paul Wessel https://www.soest.hawaii.edu/pwessel/gshhg/. Monthly NAO data was downloaded from https://www.cpc.ncep.noaa.gov/products/precip/CWlink/pna/norm.nao.monthly.b5001.current.ascii.table.

## Acknowledgements

This study was sponsored by the National Natural Science Foundation of China (Grant No.42006036).

42130409) and the Fundamental Research Funds for the Central Universities (B210202141).

## Conflict of interest

The authors declare no competing interests.

# Supporting Information for "Recent Increases in Tropical Cyclone Rapid Intensification Events in Global Coastal Regions"

Yi Li[1,2,3], Youmin Tang[3,4,5]*, Shuai Wang[6], Ralf Toumi[7], Xiangzhou Song[1,2,3]

[1]Key Laboratory of Marine Hazards Forecasting, Ministry of Natural Resources, Hohai University, Nanjing, China

[2]Key Laboratory of Ministry of Education for Coastal Disaster and Protection, Hohai University, Nanjing, China

[3]College of Oceanography, Hohai University, Nanjing, China

[4]University of Northern British Columbia, Prince George, Canada

[5]Southern Marine Science and Engineering Guangdong Laboratory (Zhuhai), Zhuhai, China

[6] Program in Atmospheric and Oceanic Sciences, Princeton University, Princeton, NJ, USA

[7]Department of Physics, Imperial College London, London, SW7 2AZ, UK

*Corresponding author: Youmin Tang, ytang@unbc.ca

**Content:**

Text S1.

Figures S1 to S11.

# Text S1. Improved ordinary least square regression

We implement an improved Ordinary Least-Square (OLS) algorithm[1] because the robustness of linear regression is usually prone to the choice of period. This OLS algorithm assumes a time series can be decomposed in the following form,

$$y(t) = Bt + \sum A_i \cdot \sin(\omega_i t + \varphi_i) + N_t, \qquad (1)$$

where B is the true secular linear trend coefficient. $\sum A_i \cdot \sin(\omega_i t + \varphi_i)$ denotes multiscale internal variations, with $A_i$, $\omega_i$ and $\varphi_i$ being the amplitude, frequency, and phase of $i$th oscillatory term, respectively. $N_t$ represents noises. As derived by Lian (2017)[1], the linear trend of Equation (1) is

$$r(L) \cong B - \frac{6}{L^2} \sum \frac{A_i}{\omega_i} [\cos(\omega_i L + \varphi_i) + \cos(\varphi_i)], \qquad (2)$$

which consists of two components, one representing the true trend and one representing the uncertainty caused by internal oscillations. As the magnitude of the sum of the terms in the bracket in Equation (2) cannot be greater than 2, we have

$$||r(L) - B|| \leq B_{th} = \frac{12}{L^2} \sum \frac{A_i}{\omega_i}, \qquad (3)$$

where $B_{th}$ is the theoretical threshold. For a given amplitude, low-frequency variations (small $\omega_i$) will have a greater impact on the estimated trend than high-frequency variations (large $\omega_i$). The secular linear trend coefficient B has the upper and lower limits of $r(L) - B_{th} \leq B \leq r(L) + B_{th}$. Then one can calculate if the sign of B would be the same as r(L) by comparing the amplitudes of r(L) and $B_{th}$.

**References**

1. Lian, T. Uncertainty in detecting trend: a new criterion and its applications to global SST. *Clim Dyn* **49**, 2881–2893 (2017).

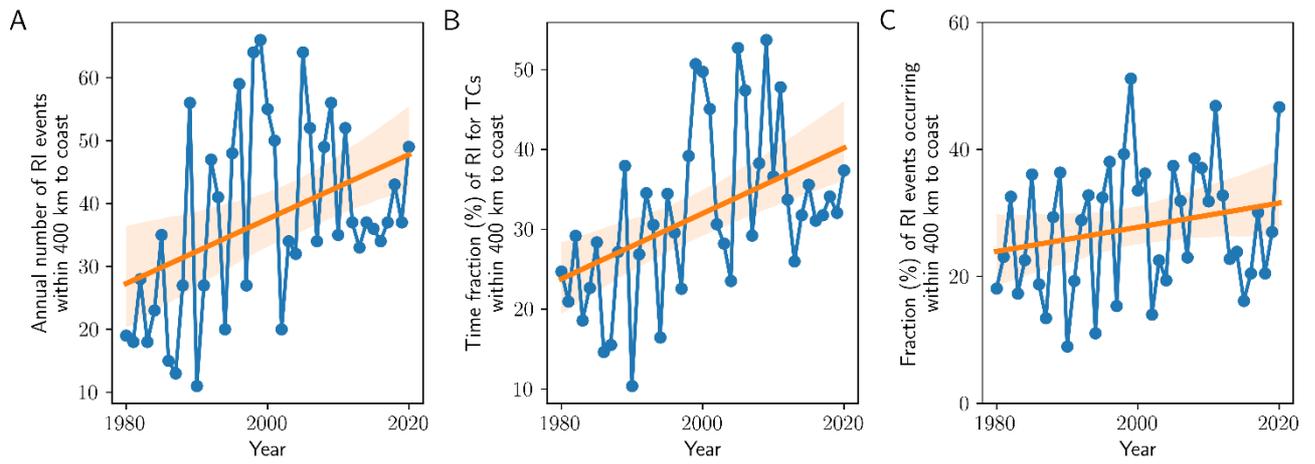

**Fig. S1. Time series of global RI activity in the coastal regions. A** Annual mean count of RI events within 400 km from the coast. **B** Annual mean time fraction of RI for TCs within 400 km from the coast. **C** Ratio of global RI events occurring within 400 km from the coast. RI is defined as an intensification of at least 30 kt/24 h. The blue lines and dots show historical data. The orange lines show linear trends, with shading areas denoting a 95% confidence interval.

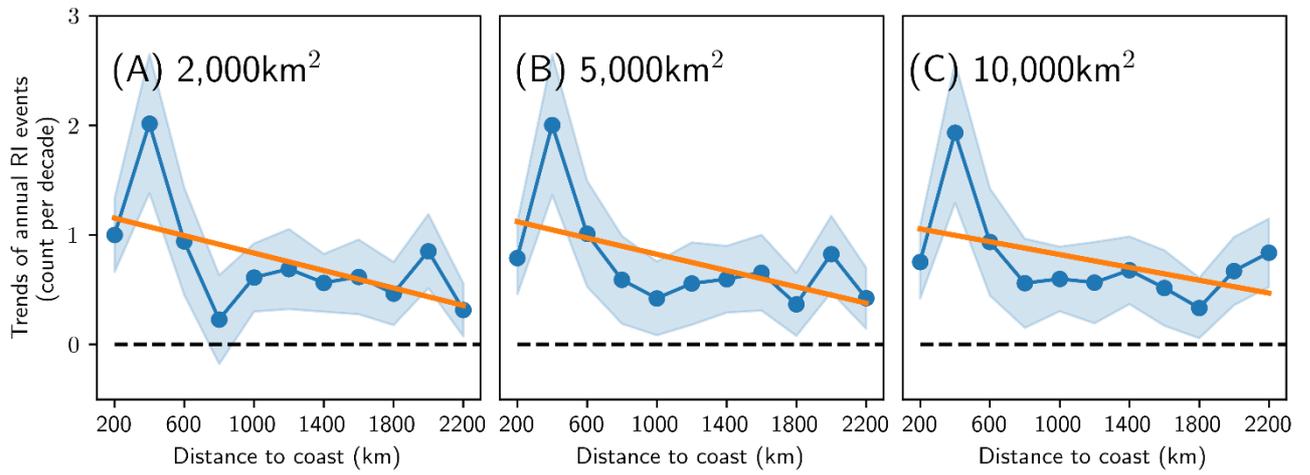

**Fig. S2. Trends of annual RI counts with different distance to land, with the minimum landmass size of (A) 2000 km², (B) 5000 km², and (C) 10000 km².** RI is defined as an intensification of at least 45 kt/24 h. The x-axis is the distance to land from 0–200 km to 2000–2200 km, with a 200-km interval. The blue lines and shadings show linear temporal trends and 95% confidence level of the trends, respectively. The orange lines show linear fits of the temporal trends as a function of distance to land.

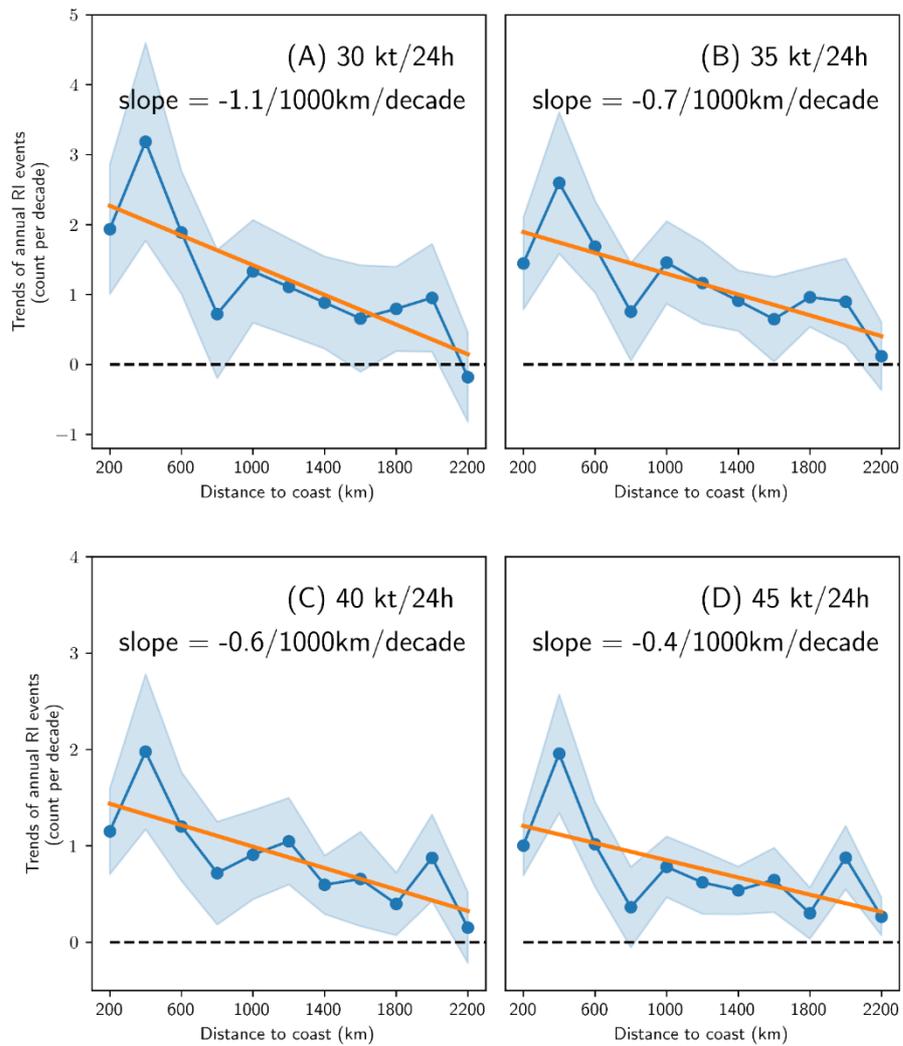

**Fig. S3. Trends of annual RI counts with different distance to land, with the RI defined as an intensification of at least (A) 30 kt/24h, (B) 35 kt/24h, (C) 40 kt/24h, and (D) 45 kt/24h**. The x-axis is the distance to land from 0-200 km, to 2000-2200 km with a 200-km interval. The blue lines and shadings show linear temporal trends and 95% confidence level of the trends, respectively. The orange lines show the linear fits of the temporal trends as a function of distance to land, with the slopes labeled in each subplot.

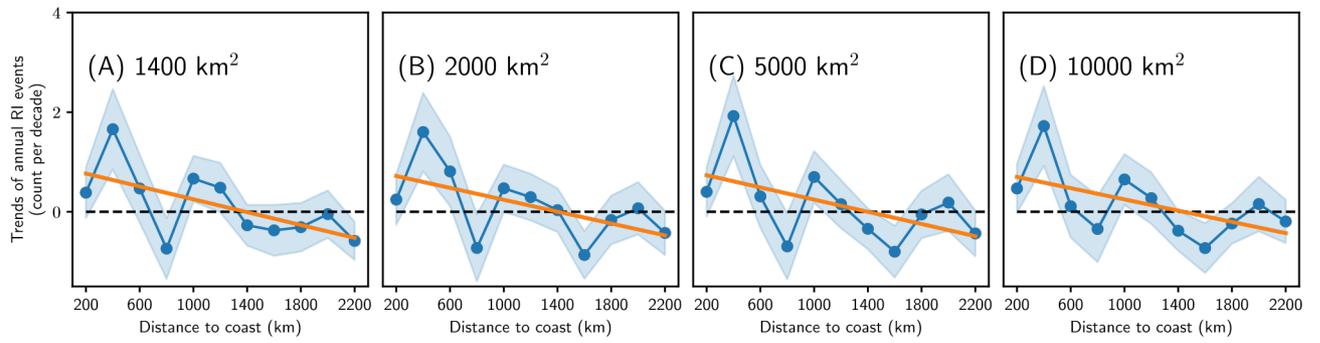

**Fig. S4. Trends of annual RI counts in ADT-HURSAT dataset with different distance to land, with the minimum landmass size of (A) 1400 km², (B) 2000 km², (C) 5000 km², and (D) 10000 km².** The RI is defined as an intensification of at least 45 kt/24 h. The x-axis is the distance to land from 0–200 km to 2000–2200 km, with a 200-km interval. The blue lines and shadings show linear temporal trends and 95% confidence level of the trends, respectively. The orange lines show linear fits of the temporal trends as a function of distance to land.

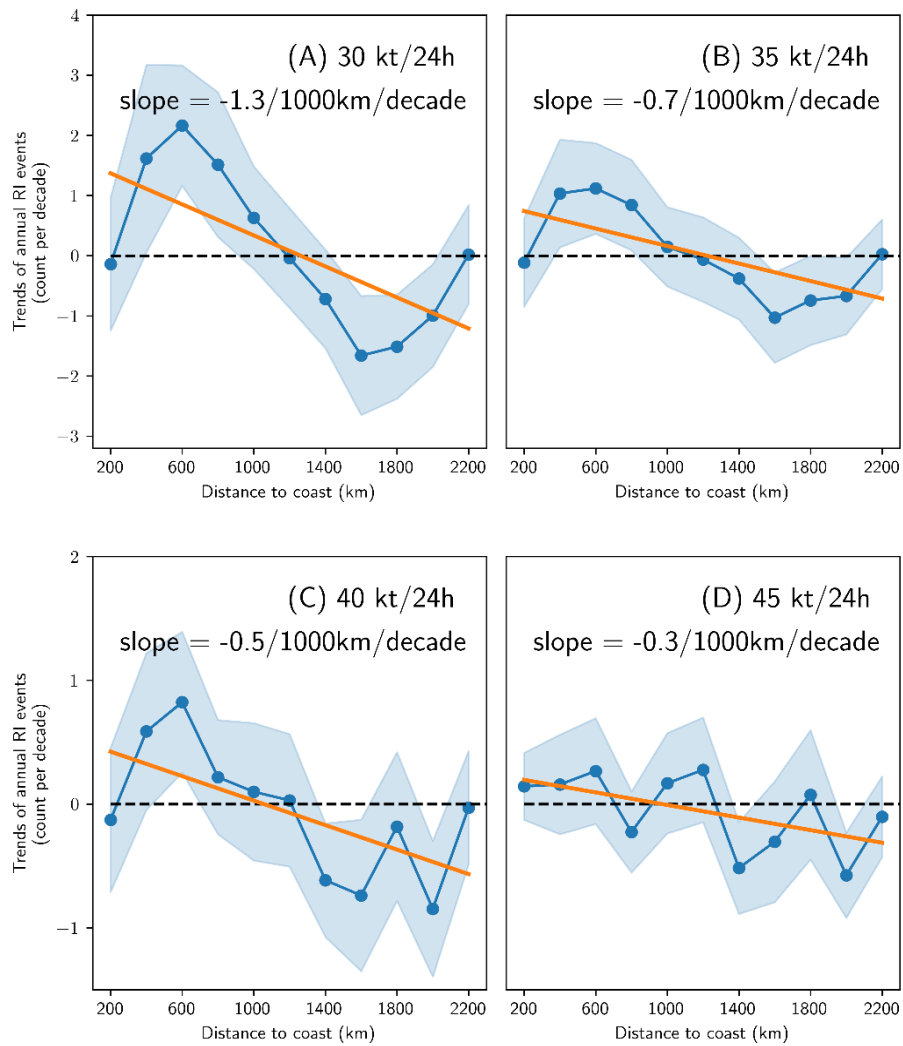

**Fig. S5. Trends of annual RI counts from 1951 to 1979 with different distance to land, with the RI defined as an intensification of at least (A) 30 kt/24h, (B) 35 kt/24h, (C) 40 kt/24h, and (D) 45 kt/24h.** The x-axis is the distance to land from 0-200 km, to 2000-2200 km with a 200-km interval. The blue lines and shadings show linear temporal trends and 95% confidence level of the trends, respectively. The orange lines show the linear fits of the temporal trends as a function of distance to land, with the slopes labeled in each subplot.

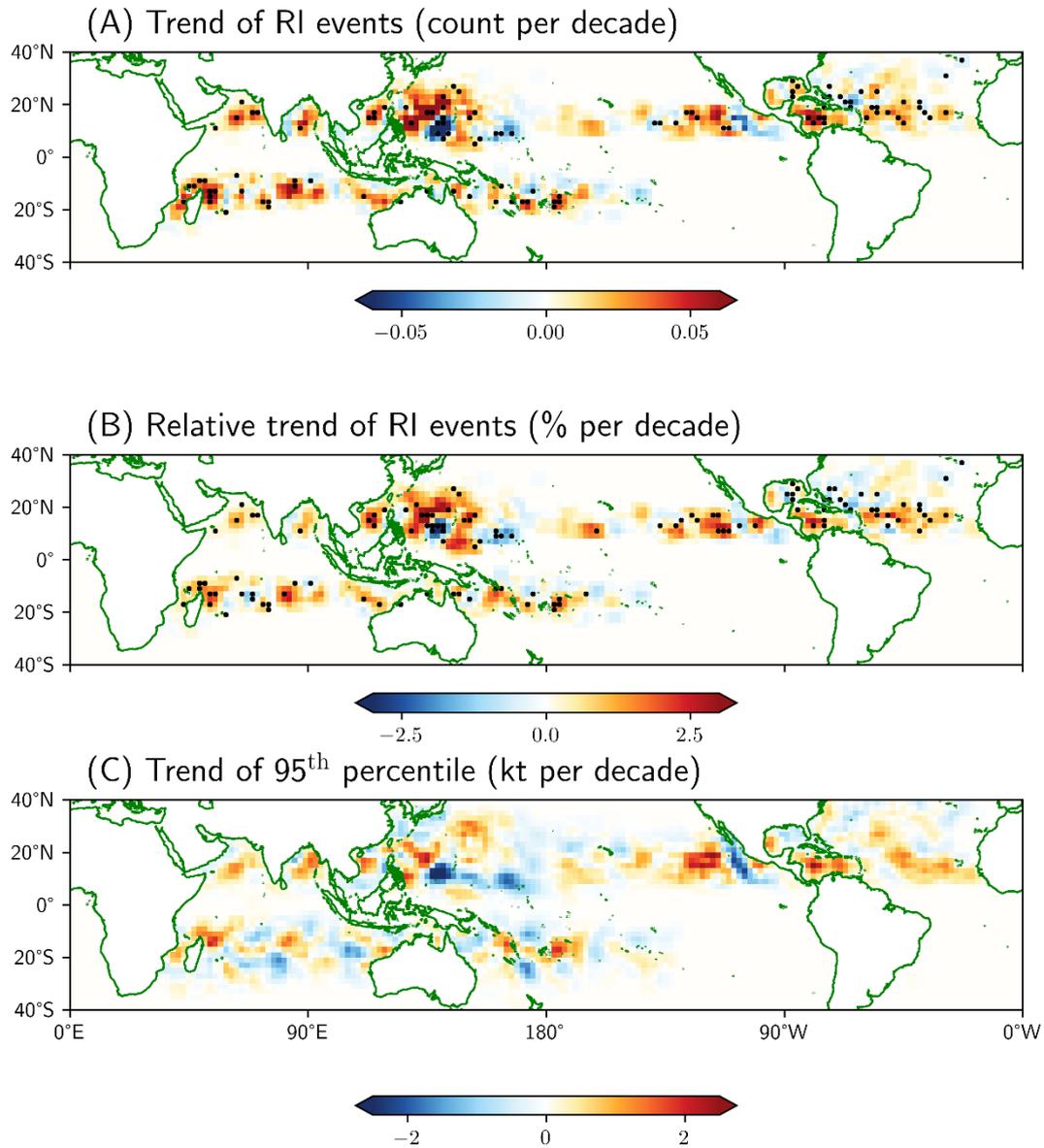

**Fig. S6. Spatial distribution of the increasing trends of annual RI counts events. A** Trend in annual count of RI events. **B** Relative trend in annual count of RI events. **C** The 95$^{th}$ percentile of 24-h intensity change. RI was defined as an intensification of at least 30 kt/24 h. The counts and ratios were calculated for each 2×2 latitude-longitude grid. Black dots show areas where 95% confidence for the linear fit was satisfied. Data smoothing using a three-point smoother was performed for better display clarity.

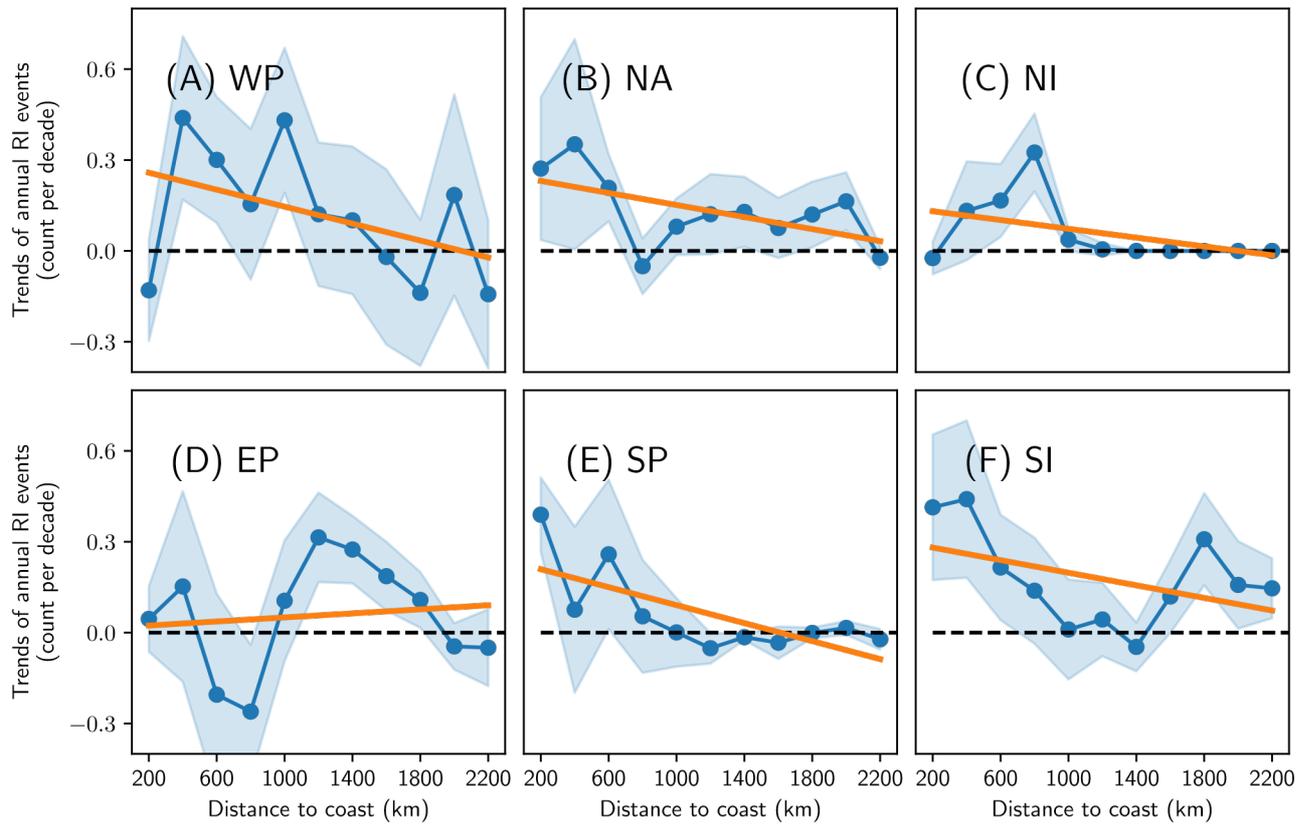

**Fig. S7. Trends of annual RI counts with different distance to land for the (A) western North Pacific (WP), (B) North Atlantic (NA), (C) North Indian Ocean (NI), (D) eastern North Pacific (EP), (E) South Pacific (SP), and (D) South Indian Ocean.** The x-axis is the distance to land from 0-200 km, to 2000-2200 km with a 200-km interval. The blue lines and shadings show linear temporal trends and 95% confidence level of the trends, respectively. The orange lines show the linear fits of the temporal trends as a function of distance to land.

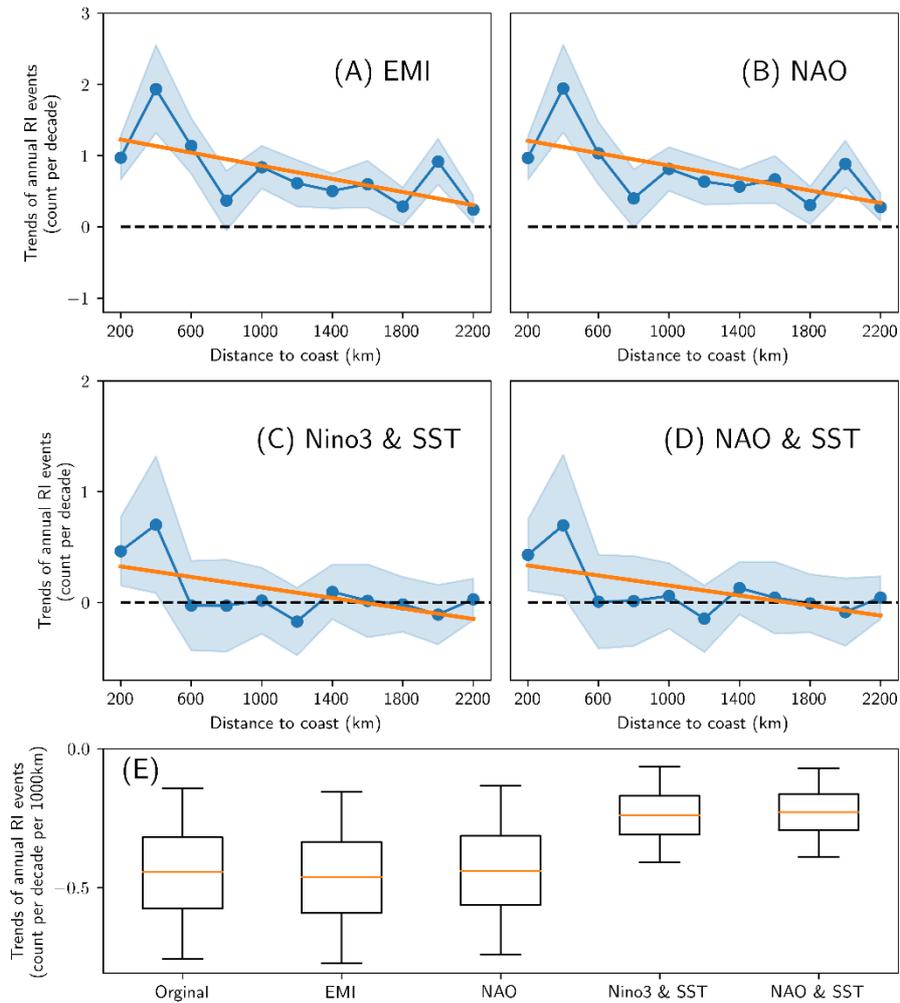

**Fig. S8. Trends of RI events with climate indices and/or global SST trend reduced. A-D** Trend of annual count of RI events with **A** Enso-Modoki index, **B** North Atlantic Oscillation, **C** both Nino3 index and global SST trend, and **D** both NAO and SST trend reduced. The x-axis is the distance to land from 0–200 km to 2000–2200 km, with a 200–km interval. The blue lines and shadings show linear temporal trends and a 95% confidence level of the trends, respectively. The orange lines show linear fits of the temporal trends as a function of distance to land. **E** Magnitudes of landward variation (i.e., slopes of orange lines in **A-D**). Orange line are the median, boxes extend from the 25th to 75th percentiles, whiskers show the 5th and 95th percentiles.

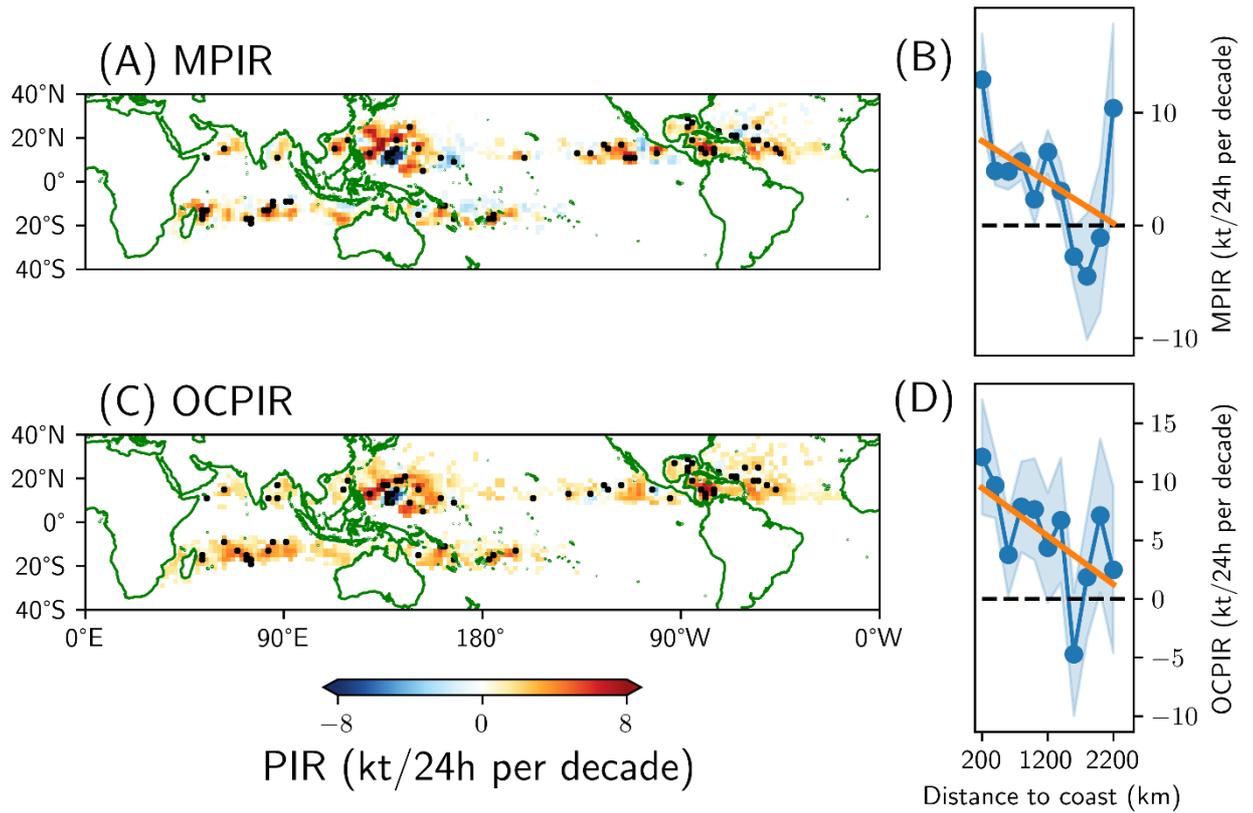

**Fig. S9. Linear trend of potential intensification rate.** **A** Spatial distribution of the linear trend of maximum potential intensification rate (MPIR, calculated from MPI). **B** Linear trend of MPI with different distances-to-land. **C** Spatial distribution of the linear trend of ocean-coupling potential intensification rate (OCPIR, calculated from OCPI). **D** Linear trend of MPI with different distances-to-land.

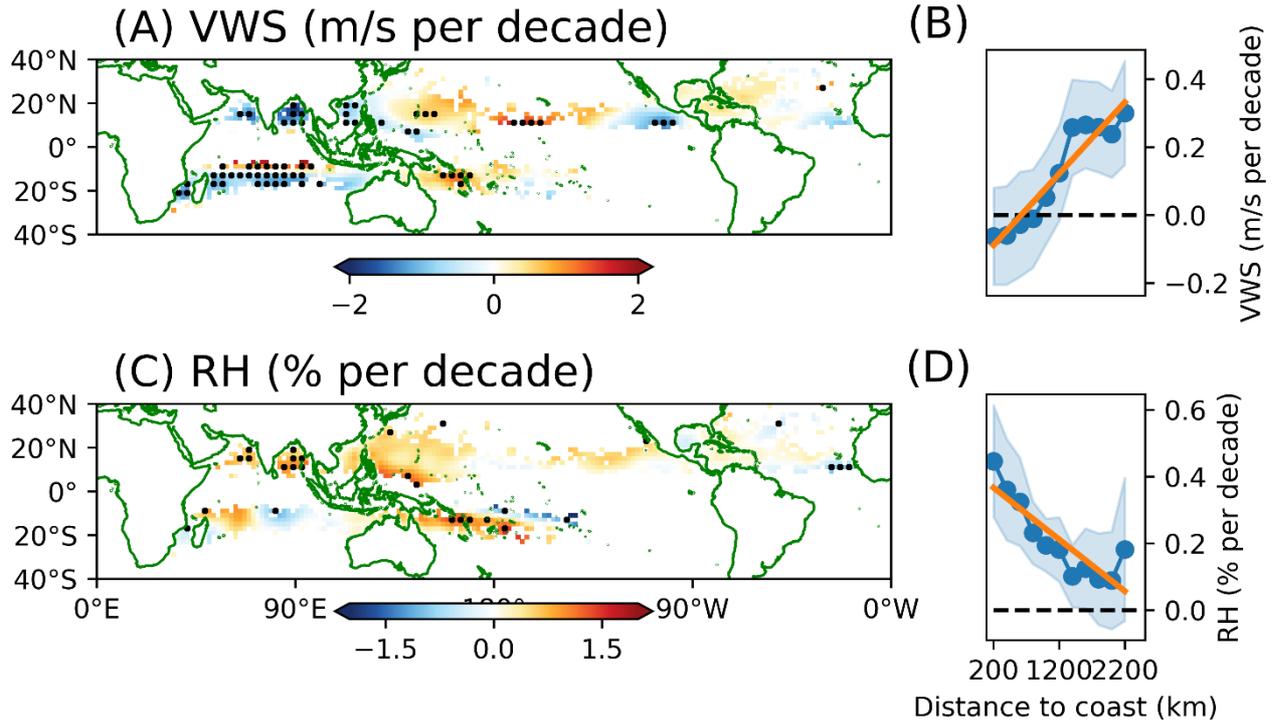

Fig. S10. Linear trend of ambient atmospheric conditions. A Spatial distribution of the linear trend of vertical wind shear (VWS). B Linear trend of VWS with different distances-to-land. C Spatial distribution of the linear trend of relative humidity (RH). D Linear trend of RH with different distances-to-land.

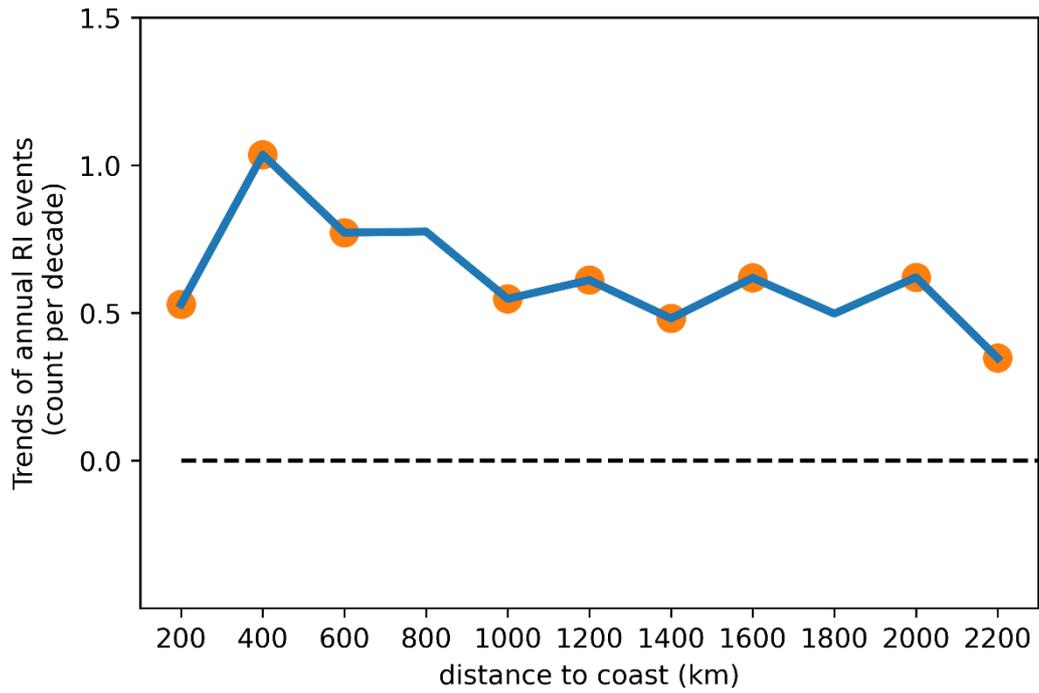

**Fig. S11. Trends of annual RI counts with different distance to land, calculated using the improved ordinary least square algorithm**[1]**.** RI is defined as an intensification of at least 45 kt/24 h. The x-axis is the distance to land from 0–200 km to 2000–2200 km, with a 200-km interval. The blue lines show linear temporal trends. The orange dots show the points where the trends are significant.